\documentclass[floats, prd, eqnum, showpacs, nofootinbib, 
twocolumn, 
eqsecnum]{revtex4-1}

\usepackage{color,graphicx}
\usepackage{amsfonts}
\usepackage{amssymb}
\usepackage{comment}

\begin{document}
    
\title{Gravitational lensing outside and inside of a marginally unstable photon sphere at a $Z_2$-symmetric wormhole throat in strong deflection limits}
\author{Naoki Tsukamoto${}^{1}$}\email{tsukamoto@rikkyo.ac.jp}
\affiliation{
${}^{1}$Department of Physics, Faculty of Science, Tokyo University of Science, 1-3, Kagurazaka, Shinjuku-ku, Tokyo 162-8601, Japan \\
}
\begin{abstract}
The deflection angles of rays diverge logarithmically near outside and inside of a photon sphere in usual strong deflection limits 
and the rays near the photon sphere could be detected by near-future space observations on black hole shadows.
On the other hand, we show that the deflection angles of the rays near the photon sphere at a throat of a $Z_2$-symmetric wormhole, 
as an example among some commonly used types of wormholes or simple wormholes,
which is often studied by the Event Horizon Telescope collaborations, diverge in power law but not logarithmically.
This is because the $Z_2$-symmetric wormhole has a marginally unstable photon sphere at the throat and it has neither a common nor simple but special property. 
We investigate gravitational lensing of the rays nearly outside and inside of the marginally unstable photon sphere at the throat in a class of $Z_2$-symmetric wormholes in the strong deflection limits.
We use a numerical method to calculate the deflection angles diverging in power law and we apply it 
to the commonly-used simple wormhole spacetime, a Damour-Solodukhin wormhole spacetime, and a Simpson-Visser black-bounce spacetime.
Our numerical results imply universal property of the deflection angle of the rays near the marginally unstable photon sphere on $Z_2$-symmetric wormhole throat in the strong deflection limits
as similar to the deflection angles of ray bent by a marginally unstable photon sphere off a throat in the strong deflection limits which have a different power.
We also correct the coefficients of power-law divergent terms of the deflection angles of the rays bent nearly outside of the marginally unstable photon sphere
at the $Z_2$-symmetric wormhole throat in a semi-analytic approach investigated by the author and others.  
\end{abstract}

\maketitle

\section{Introduction}
Recently, LIGO Scientific and VIRGO Collaborations have reported the direct detection of gravitational waves from binary black holes~\cite{Abbott:2016blz,LIGOScientific:2026wfs}
and Event Horizon Telescope Collaboration has reported ring images around supermassive black hole candidates at the centers of the giant elliptical galaxy M87~\cite{Akiyama:2019cqa} 
and the Milky Way Galaxy~\cite{EventHorizonTelescope:2022wkp}.

Black holes and other compact objects such as wormholes with strong gravity make photon spheres which are a set of unstable circular light orbits~\cite{Perlick_2004_Living_Rev}
and the radii of the photon spheres~\cite{Hod:2017xkz,Peng:2018nkj,Hod:2020pim,Hod:2023jmx}, their numbers~\cite{Hod:2017zpi,Cunha:2020azh,Cunha:2024ajc,Padhye:2024yrj}, 
and relationships between the photon spheres and photon absorption cross sections~\cite{Sanchez:1977si,Decanini:2010fz,Wei:2011zw},
quasinormal modes~\cite{Press:1971wr,Goebel_1972,Stefanov:2010xz,Raffaelli:2014ola,Igata:2025plb,Igata:2025hpy},
a centrifugal force and gyroscopic precession~\cite{Abramowicz_Prasanna_1990,Abramowicz:1990cb,Allen:1990ci,Hasse_Perlick_2002},
Bondi's sonic horizon~\cite{Mach:2013gia,Chaverra:2015bya,Cvetic:2016bxi,Koga:2016jjq,Koga:2018ybs,Koga:2019teu},
the stability of thin-shell wormholes~\cite{Barcelo:2000ta,Koga:2020gqd,Tsukamoto:2023kvk},
gravitational lenses~\cite{Hagihara_1931,Darwin_1959,Atkinson_1965,Luminet_1979,Ohanian_1987,Nemiroff_1993,Frittelli_Kling_Newman_2000,Virbhadra_Ellis_2000,Bozza:2001xd,Eiroa:2002mk,Bozza:2002zj,Bozza:2002af,Perlick:2003vg,Nandi:2006ds,Virbhadra:2008ws,Hioki:2009na,Bozza_2010,Tsupko:2017rdo,Nakao:2018knn,Gralla:2019xty,Okabayashi:2020apz,Hsieh:2021scb,Aratore:2021usi,Tsupko:2022kwi,Aratore:2024bro,Feleppa:2024vdk,Feleppa:2025ejh,Tsupko:2025hhf}, 
and the apparent shape of collapsing stars~\cite{Ames_1968,Synge:1966okc,Yoshino:2019qsh,Koga:2025dii}
were considered and
alternatives to the photon spheres were also suggested in Refs.~\cite{Claudel:2000yi,Koga:2019uqd,Cunha:2017eoe,Gibbons:2016isj,Shiromizu:2017ego,Yoshino:2017gqv,Galtsov:2019bty,Galtsov:2019fzq,Siino:2019vxh,Yoshino:2019dty,Cao:2019vlu,Yoshino:2019mqw,Lee:2020pre}. 
Ultracompact objects with an antiphoton sphere, which is a set of stable circular light orbits,
may be unstable due to the slow decay of linear waves near the antiphoton sphere~\cite{Keir:2014oka,Cardoso:2014sna,Cunha:2017qtt,Cunha:2017eoe,Cunha:2022gde,Zhong:2022jke}.     

The rays bent near the photon spheres of the supermassive black hole candidates at the centers of M87 and the Milky Way Galaxy 
could be detected by near-future space observations~\cite{Lupsasca:2024xhq,Johnson:2024ttr}   
while they are not detected by the Event Horizon Telescope~\cite{Akiyama:2019cqa,EventHorizonTelescope:2022wkp}. 
The near-future space observations of their ring images~\cite{Lupsasca:2024xhq,Johnson:2024ttr} 
and the future observations of S301~\cite{Dayem:2026ktt,Piran:2026zjs} will determine 
the spin of Sagittarius A* which is the supermassive black hole candidate at the center of the Milky Way Galaxy.   
To exclude black hole mimickers with the photon spheres from the near-future observations 
we should investigate the details of the photon spheres of exotic compact objects such as wormholes~\cite{Ellis:1973yv,Chetouani:1984qdm,Perlick:2003vg,Muller:2004dq,Nandi:2006ds,Muller:2008zza,Tsukamoto:2012xs,Ohgami:2015nra,Tsukamoto:2016qro,Tsukamoto:2016zdu,Tsukamoto:2017edq,Shaikh:2018kfv,Shaikh:2018oul,Bronnikov:2018nub,Shaikh:2019jfr,Wielgus:2020uqz,Bronnikov:2021liv,Maeda:2021wnl,Olmo:2023lil,Bugaev:2023mlc,Zhang:2024sgs,Xavier:2024iwr,Tsukamoto:2024pid,Bugaev:2024dte,Solodukhin:2025opw} 
and naked singularity~\cite{Virbhadra:1998dy,Virbhadra:2002ju,Hioki:2009na,Shaikh:2018lcc,Shaikh:2019itn}.  

In 2002, Eiroa, Romero, and Torres~\cite{Eiroa:2002mk} and Bozza~\cite{Bozza:2002zj} investigated numerical and semi-analytic methods, respectively,
to calculate the deflection angle $\alpha$ of rays reflected by just outside of the photon sphere in a general, static, spherically symmetric, and asymptotically-flat spacetime 
in a strong deflection limit
$b \rightarrow b_{\mathrm{m}} + 0$, where $b$ and $b_{\mathrm{m}}$ are the impact parameter and the critical impact parameter of the rays, 
and they considered gravitational lensing~\cite{Schneider_Ehlers_Falco_1992,Schneider_Kochanek_Wambsganss_2006} by the photon sphere.
Their methods and extended methods were studied eagerly~\cite{Bozza:2002af,Eiroa:2002mk,Petters:2002fa,Eiroa:2003jf,Bozza:2004kq,Bozza:2005tg,Bozza:2006sn,Bozza:2006nm,Iyer:2006cn,Bozza:2007gt,Tsukamoto:2016qro,Ishihara:2016sfv,Tsukamoto:2016oca,Tsukamoto:2016zdu,Tsukamoto:2016jzh,Shaikh:2018oul,Shaikh:2019itn,Takizawa:2021gdp,Zhang:2024sgs,Igata:2025taz}.
The deflection angle of the rays reflected by the photon sphere in strong deflection limits~$b \rightarrow b_{\mathrm{m}} \pm 0$~\cite{Shaikh:2019itn,Shaikh:2019jfr,Tsukamoto:2021fsz} 
can be expressed by 
\begin{eqnarray}\label{eq:def0}
\alpha&=&-\bar{a}_\pm \log \left| \frac{b}{b_{\mathrm{m}}}-1 \right| + \bar{b}_\pm,
\end{eqnarray}
where the upper and lower signs of the coefficients $\bar{a}_\pm$ and the terms $\bar{b}_\pm$ 
are chosen for the rays outside and inside of the photon sphere, respectively. 

If a photon sphere and an antiphoton sphere degenerate to be a marginally unstable photon sphere, 
the deflection angles $\alpha$ of the rays in the strong deflection limits~$b \rightarrow b_{\mathrm{m}} \pm 0$ diverge in power law in the following form:
\begin{equation}\label{eq:def01}
\alpha(b)=\frac{\bar{c}_\pm}{\left| \frac{b}{b_{\mathrm{m}}}-1\right|^\frac{1}{6} } +\bar{d}_\pm, 
\end{equation}
where the upper and lower signs of the coefficients $\bar{c}_\pm$ and the terms $\bar{d}_\pm$ are chosen for for the rays reflected near outside and inside of the marginally unstable photon sphere. 
The spacetimes can be interested to know the most different cases from the Schwarzschild black hole among black-hole mimickers with (marginally unstable) photon spheres.   
Chiba and Kimura~\cite{Chiba:2017nml} obtained a coefficient correspond to $\bar{c}_+$ in a Hayward spacetime~\cite{Hayward:2005gi} but not the constant $\bar{d}_+$.
In Ref.~\cite{Tsukamoto:2020iez}, Tsukamoto suggested a semi-analytic formula to obtain $\bar{c}_+$ and $\bar{d}_+$ in a general, static, spherically symmetric, and asymptotically-flat spacetime 
with a marginally unstable photon sphere and obtained $\bar{c}_+$ and $\bar{d}_+$ in the Reissner-Nordstr\"{o}m spacetime and the Hayward spacetime. 
Sasaki~\cite{Sasaki:2025web} obtained the anatific forms of $\bar{c}_\pm$ and $\bar{d}_\pm$ in the Reissner-Nordstr\"{o}m spacetime 
and pointed out that $\bar{d}_+$ is correct but $\bar{c}_+$ is not in Tsukamoto~\cite{Tsukamoto:2020iez}.
In Ref.~\cite{Tsukamoto:2025hbz}, Tsukamoto extended Eiroa, Romero, and Torres' method~\cite{Eiroa:2002mk} to 
the general, static, spherically symmetric, and asymptotically-flat spacetime with the marginally unstable photon sphere in strong deflection limits~$b \rightarrow b_{\mathrm{m}} \pm 0$
and numerically obtained correct coefficients $\bar{c}_\pm$ and the terms $\bar{d}_\pm$ in the Reissner-Nordstr\"{o}m and Hayward spacetimes.
Igata, Sasaki, and Tsukamoto \cite{Igata:2026ivq} obtained $\bar{c}_\pm$ and $\bar{d}_\pm$ analytically 
in a general, static, spherically symmetric, and asymptotically-flat spacetime with the marginally unstable photon sphere
and applied to the Reissner-Nordstr\"{o}m, Hayward, Bardeen, and Reissner-Nordstr\"{o}m-like wormhole spacetimes. 
They found universal relations $\bar{c}_- = \sqrt{3} \bar{c}_+$ and $\bar{d}_- = \bar{d}_+$. 

The Event Horizons Telescope Collaboration claimed that
a $Z_2$-symmetric wormhole spacetime~\cite{Harko:2008vy,Bambi:2013jda,Bambi:2013nla}, 
which they considered as an example among some commonly used types of wormholes or one of the simple wormholes, 
is incompatible with their observations of the centers of M87 and the Milky Way Galaxy~\cite{Akiyama:2019cqa,EventHorizonTelescope:2022xqj}.
In this paper, however, we show that the $Z_2$-symmetric wormhole has a marginally unstable photon sphere at a throat
and that it causes neither a common nor simple but special property.  

In this paper, we extend Eiroa, Romero, and Torres' method~\cite{Eiroa:2002mk} to 
the general, static, spherically symmetric, asymptotically-flat $Z_2$-symmetric wormhole spacetime with the marginally unstable photon sphere at the throat 
in the strong deflection limits~$b \rightarrow b_{\mathrm{m}} \pm 0$.
The deflection angle of the rays, which are emitted by a light source at our side and another side against the $Z_2$-symmetrical wormhole throat 
and which deflected just outside and inside, respectively, of the marginally unstable photon sphere on the throat, 
can be expressed by~\cite{Tsukamoto:2020uay,Tsukamoto:2020bjm,Zhang:2024sgs}
\begin{equation}\label{eq:def02}
\alpha(b)=\frac{\bar{c}_\pm}{\left| \frac{b}{b_{\mathrm{m}}}-1\right|^\frac{1}{4} } +\bar{d}_\pm
\end{equation}
in the strong deflection limits~$b \rightarrow b_{\mathrm{m}} \pm 0$.
We apply our method to the commonly-used simple wormhole spacetime~\cite{Harko:2008vy,Bambi:2013jda,Bambi:2013nla}, 
a Damour-Solodukhin wormhole spacetime~\cite{Damour:2007ap} 
and a Simpson-Visser black-bounce spacetime~\cite{Simpson:2018tsi}
with the marginally unstable photon sphere at the throat.
Our numerical results imply that universal relations $\bar{c}_- = \sqrt{2} \bar{c}_+$ and $\bar{d}_- = \bar{d}_+$.

We also show that the coefficient $\bar{c}_+$ is invalid while the terms $\bar{d}_+$ is correct in the Damour-Solodukhin wormhole spacetime in Tsukamoto \cite{Tsukamoto:2020uay}.
This implies that a semi-analytic method of gravitational lensing of 
the rays just outside of the marginally unstable photon sphere on the throat in general, static, spherically symmetric, and asymptotically-flat wormhole spacetimes 
with the $Z_2$ symmetry against the throat in the strong deflection limit~$b \rightarrow b_{\mathrm{m}} + 0$ 
suggested by Zhang and Xie~\cite{Zhang:2024sgs} also gives correct terms $\bar{d}_+$ and the invalid coefficient $\bar{c}_+$
since they generalized the method in Ref.~\cite{Tsukamoto:2020uay}. 

This paper is organized as follows. 
In Sec.~II, we investigate the deflection angle of the rays reflected just outside and inside of the marginally unstable photon sphere at 
the throat in the general, static, spherically symmetric, asymptotically-flat $Z_2$-symmetric wormhole spacetime  
in the strong deflection limits~$b \rightarrow b_{\mathrm{m}} \pm 0$.
In Sec.~III, we solve a lens equation.
In Sec.~IV, we apply our method to the commonly-used simple wormhole spacetime~\cite{Harko:2008vy,Bambi:2013jda,Bambi:2013nla}, the Damour-Solodukhin wormhole spacetime~\cite{Damour:2007ap} 
and the Simpson-Visser black-bounce spacetime~\cite{Simpson:2018tsi,Nascimento:2020ime}.
In Sec.~V, we conclude and discuss our results.
We use a unit that a light speed and Newton's constant are unity in this paper.

\section{Deflection angle of rays by a wormhole with $Z_2$ symmetry}
We consider a general, static, spherically symmetric, asymptotically flat, wormhole spacetime.
Its line element can be expressed by, in coordinates $x^\mu=(t, r, \vartheta, \varphi)$,
\begin{eqnarray}\label{eq:LE0}
ds^2
=-A(r)\mathrm{d}t^2+ B(r)\mathrm{d}r^2 +C(r) (\mathrm{d}\vartheta^2+\sin^2 \vartheta \mathrm{d}\varphi^2), \nonumber\\
\end{eqnarray}
where the coordinates are defined in domains $-\infty <t< \infty, -\infty <r< \infty, 0 \leq \vartheta \leq \pi$, and $0 \leq \varphi <2 \pi$,
and we assume that $A(r)$, $B(r)$, and $C(r)$ are positive and finite in the domains and
that the wormhole has $Z_2$ symmetry against its throat at $r=r_\mathrm{th}$.
The functions $A(r)$, $B(r)$, and $C(r)$ satisfy asymptotically-flat conditions 
\begin{eqnarray}\label{eq:asymptotically-flat-conditions1} 
&&\lim_{r \rightarrow \pm \infty} A(r) = \lim_{r \rightarrow \pm \infty} B(r) = 1+O(r^{-1}), \\ \label{eq:asymptotically-flat-conditions2} 
&&\lim_{r \rightarrow \pm \infty} C(r) = r^2+O(r).
\end{eqnarray}
The wormhole spacetime has time-translational and axial Killing vectors $t^\mu \partial_\mu = \partial_t$ and $\varphi^\mu \partial_\mu = \partial_\varphi$ 
because of its stationarity and axial symmetry, respectively.
We assume that rays are on an equatorial plane $\vartheta =\pi/2$ without loss of generality due to its spherical symmetry.

The line elements of static, spherically symmetric, asymptotically flat, and $Z_2$-symmetric wormhole spacetimes are often described by, in coordinates $(t, \rho, \vartheta, \varphi)$,
\begin{eqnarray}\label{eq:LE1}
ds^2
=-A(\rho)\mathrm{d}t^2+ g_{\rho \rho}(\rho)\mathrm{d}\rho^2 +C(\rho) (\mathrm{d}\vartheta^2+\sin^2 \vartheta \mathrm{d}\varphi^2), \nonumber\\
\end{eqnarray}
where the radial coordinate $\rho$ is defined in a domain $\rho_\mathrm{th} \leq \rho < \infty$, 
the metric component $g_{\rho \rho}(\rho)$ diverges at the throat $\rho=\rho_\mathrm{th}$, 
it holds $\mathrm{d}g_{\rho \rho}/\mathrm{d} \rho \leq 0$ near the throat.
We assume $g_{\rho \rho}(\rho)>0$ in the domain $\rho_\mathrm{th} \leq \rho < \infty$.
By using 
\begin{eqnarray}
r=\int^r_0 \mathrm{d}r= \pm \int^\rho_{\rho_\mathrm{th}} \sqrt{g_{\rho \rho}} \mathrm{d}\rho,
\end{eqnarray}
the line element~(\ref{eq:LE1}) is rewritten in
\begin{eqnarray}\label{eq:LE2}
ds^2
=-A(\rho(r))\mathrm{d}t^2+ \mathrm{d}r^2 +C(\rho(r)) (\mathrm{d}\vartheta^2+\sin^2 \vartheta \mathrm{d}\varphi^2).  \nonumber\\
\end{eqnarray}
Here we have set that the throat is at $r=r_\mathrm{th} \equiv 0$ and that the metric function $B(r)$ is unity.
The function $C(r)$ must hold relations $C^{\prime}(0)=0$ at the throat $r=r_\mathrm{th} = 0$ and $C^{\prime \prime}(r)>0$ near the throat,
where $\prime$ denotes a differentiation with respect to $r$.

From $k^\mu k_\mu=0$, where $k^\mu$ is the wave number of a ray defined by $k^\mu \equiv \dot{x}^\mu$, where the dot denotes a differentiation with respect to an affine parameter along the ray, 
the trajectory of the ray is given by 
\begin{eqnarray}\label{eq:trajectory1}
-A(r)\dot{t}^2+B(r)\dot{r}^2+C(r)\dot{\varphi}^2=0
\end{eqnarray}
and, by rescaling the affine parameter, it can be expressed by    
\begin{eqnarray}
\dot{r}^2+V(r,b)=0,
\end{eqnarray}
where $V(r,b)$ is the effective potential of the ray for its radial motion defined as 
\begin{eqnarray}
V(r,b) \equiv \frac{1}{B(r)} \left( \frac{b^2}{C(r)} -\frac{1}{A(r)} \right),
\end{eqnarray}
where $b \equiv L/E$ is the impact parameter of the ray and $E \equiv -g_{\mu \nu} t^\mu k^\nu=A(r)\dot{t}$ and $L=g_{\mu \nu} \varphi^\mu k^\nu=C(r)\dot{\varphi}$ are the conserved energy and angular momentum of the ray, respectively.
The ray can be in regions where the effective potential $V(r,b)$ is non-positive or $V(r,b) \leq 0$. 
From Eqs.~(\ref{eq:asymptotically-flat-conditions1}) and (\ref{eq:asymptotically-flat-conditions2}),
we obtain 
\begin{eqnarray}\label{eq:asymptotically-flat} 
&&\lim_{r \rightarrow \pm \infty} V(r,b) = -1 <0.
\end{eqnarray}
Therefore, the ray can be in two asymptotically-flat regions. 

We note that the trajectory of the ray (\ref{eq:trajectory1}) can be rewritten in 
\begin{equation}\label{eq:trajectory2}
\left( \frac{\mathrm{d}r}{\mathrm{d}\varphi} \right)^2 = \frac{C(r)}{B(r)} \left( \frac{C(r)}{A(r) b^2} -1 \right) = - \frac{C(r)^2}{b^2} V(r,b).
\end{equation}

We define 
a function $D(r)$ by 
\begin{eqnarray}
D(r)=\frac{C^{\prime}(r)}{C(r)}-\frac{A^{\prime}(r)}{A(r)}
\end{eqnarray}
and its derivatives with respect to $r$ are given by
\begin{eqnarray}
D^{\prime}(r)
&=&\frac{C^{\prime \prime}(r)}{C(r)}-\frac{C^{\prime}(r)^2}{C(r)^2}\nonumber\\
&&-\frac{A^{\prime \prime}(r)}{A(r)}+\frac{A^{\prime}(r)^2}{A(r)^2},
\end{eqnarray}
\begin{eqnarray}
D^{\prime \prime}(r)
&=&\frac{C^{\prime \prime \prime}(r)}{C(r)}-3\frac{C^{\prime \prime}(r) C^{\prime}(r)}{C(r)^2}+2\frac{C^{\prime}(r)^3}{C(r)^3}\nonumber\\
&&-\frac{A^{\prime \prime \prime}(r)}{A(r)}+3\frac{A^{\prime \prime}(r) A^{\prime}(r)}{A(r)^2}-2\frac{A^{\prime}(r)^3}{A(r)^3},
\end{eqnarray}
and
\begin{eqnarray}
D^{\prime \prime \prime}(r)
&=&\frac{C^{\prime \prime \prime \prime}(r)}{C(r)}-4\frac{C^{\prime \prime \prime}(r) C^{\prime}(r)}{C(r)^2} -3\frac{C^{\prime \prime}(r)^2}{C(r)^2}\nonumber\\
&&+12\frac{C^{\prime \prime}(r) C^{\prime}(r)^2}{C(r)^3}-6\frac{C^{\prime}(r)^4}{C(r)^4}\nonumber\\
&&-\frac{A^{\prime \prime \prime \prime}(r)}{A(r)}+4\frac{A^{\prime \prime \prime}(r) A^{\prime}(r)}{A(r)^2} +3\frac{A^{\prime \prime}(r)^2}{A(r)^2}\nonumber\\
&&-12\frac{A^{\prime \prime}(r) A^{\prime}(r)^2}{A(r)^3}+6\frac{A^{\prime}(r)^4}{A(r)^4}.
\end{eqnarray}
We also assume the wormhole has a marginally unstable photon sphere at $r=r_\mathrm{m}$
and its position corresponds to the throat at $r=r_\mathrm{th}$.
For the assumptions, we require that conditions 
\begin{eqnarray}
r_\mathrm{m}=r_\mathrm{th},
\end{eqnarray}
\begin{eqnarray}\label{eq:Dm0}
D_\mathrm{m}=\frac{C^{\prime}_\mathrm{m}}{C_\mathrm{m}}-\frac{A^{\prime}_\mathrm{m}}{A_\mathrm{m}}=0,
\end{eqnarray}
\begin{eqnarray}\label{eq:Dm1}
D^{\prime}_\mathrm{m}=\frac{C^{\prime \prime}_\mathrm{m}}{C_\mathrm{m}}-\frac{A^{\prime \prime}_\mathrm{m}}{A_\mathrm{m}}=0,
\end{eqnarray}
\begin{eqnarray}\label{eq:Dm2}
D^{\prime \prime}_\mathrm{m}=\frac{C^{\prime \prime \prime}_\mathrm{m}}{C_\mathrm{m}}-\frac{A^{\prime \prime \prime}_\mathrm{m}}{A_\mathrm{m}}=0,
\end{eqnarray}
and 
\begin{eqnarray}\label{eq:Dm3}
D^{\prime \prime \prime}_\mathrm{m}=\frac{C^{\prime \prime \prime \prime}_\mathrm{m}}{C_\mathrm{m}}-\frac{A^{\prime \prime \prime \prime}_\mathrm{m}}{A_\mathrm{m}}>0
\end{eqnarray}
should hold.
Here and hereinafter, functions with the subscript m denote the functions at $r=r_\mathrm{m}$.
\footnote{
Gravitational lensing reflected by a marginally unstable photon sphere at $r=r_\mathrm{m} \neq r_\mathrm{th}$ off the throat in wormhole specetimes or non-wormhole spacetimes, which hold conditions 
\begin{eqnarray}
D_\mathrm{m}=D^{\prime}_\mathrm{m}=0
\end{eqnarray}
and 
\begin{eqnarray}
D^{\prime \prime}_\mathrm{m}>0, 
\end{eqnarray}
were considered in Refs.~\cite{Tsukamoto:2020uay,Tsukamoto:2020bjm,Zhang:2024sgs}. 
}

Under the assumptions, the effective potential for the marginally unstable photon sphere on the throat and its derivatives satisfy conditions
\begin{equation}
V(r_\mathrm{m},b_\mathrm{m})=V^{\prime}(r_\mathrm{m},b_\mathrm{m}) 
=V^{\prime \prime}(r_\mathrm{m},b_\mathrm{m}) =V^{\prime \prime \prime}(r_\mathrm{m},b_\mathrm{m})=0,
\end{equation}
and
\begin{equation}
V^{\prime \prime \prime \prime}(r_\mathrm{m},b_\mathrm{m})
=-\frac{b_\mathrm{m}^2 D^{\prime \prime \prime}_\mathrm{m}}{B_\mathrm{m}C_\mathrm{m}}
=-\frac{D^{\prime \prime \prime}_\mathrm{m}}{A_\mathrm{m}B_\mathrm{m}}
<0,
\end{equation}
where $b_\mathrm{m}$ is the critical impact parameter of the ray defined by 
\begin{eqnarray}
b_\mathrm{m} \equiv \sqrt{\frac{C_\mathrm{m}}{A_\mathrm{m}}}.
\end{eqnarray}

We consider that a ray comes from one of the asymptotically-flat regions and that the ray passing or not passing through the throat.
If the impact parameters of the ray satisfies $\left| b \right| \geq b_\mathrm{m}$,
the ray does not pass through the throat and it goes back to the same spatial infinity. 
On the other hand, if the impact parameter $\left| b \right|$ of the ray is smaller than the critical impact parameter $b_\mathrm{m}$ or $\left| b \right| < b_\mathrm{m}$, 
the ray passes through the throat and it reaches into another spatial infinity or another asymptotically-flat region.

\subsection{Ray not passing through the throat}
In the case of $\left| b \right| \geq b_\mathrm{m}$ that the ray does not pass through the throat, 
the trajectory of the ray has the closest distance $r=r_0$ where a relation
\begin{eqnarray}\label{eq:closest1}
\dot{r}_0 \equiv \left. \dot{r} \right|_{r=r_0}=0 
\end{eqnarray}
holds and, from Eq.~(\ref{eq:trajectory1}), a relation
\begin{eqnarray}\label{eq:trajectory3}
-A_0\dot{t}_0^2+C_0\dot{\varphi}_0^2=0,
\end{eqnarray}
holds there. Here and hereinafter, functions with subscript $0$ denote the function at the closest distance $r=r_0$. 
The impact parameter $b$ can be expressed as the function of the closest distance $r_0$ as 
\begin{eqnarray}\label{eq:b0}
b=b(r_0)= \frac{L}{E} = \frac{C_0 \dot{\varphi}_0}{A_0 \dot{t}_0} = \pm \sqrt{\frac{C_0}{A_0}}.
\end{eqnarray}

From~Eqs.~(\ref{eq:trajectory2}) and (\ref{eq:b0}), 
we obtain the deflection angle $\alpha(r_0)$ of the ray as 
\begin{eqnarray}\label{eq:deflection_angle1}
\alpha(r_0) 
&\equiv& 2 \int^\infty_{r_0} \frac{\mathrm{d}r}{\sqrt{\frac{C(r)}{B(r)}\left( \frac{C(r)}{A(r)b(r_0)^2} -1 \right)}} -\pi \nonumber\\
&=& 2 \int^\infty_{r_0} \frac{ \left| b(r_0) \right| \mathrm{d}r}{C(r) \sqrt{-V(r,b(r_0))}} -\pi.
\end{eqnarray}
For simplicity, unless otherwise stated, we will assume $b>0$ below. 

The leading and sub-leading terms of the deflection angle of the ray near outside of 
the marginally unstable photon sphere on the throat with the $Z_2$ symmetry in the strong deflection limit 
$r_0 \rightarrow r_{\mathrm{m}}+0$, $\rho_0 \rightarrow \rho_{\mathrm{m}}+0$, 
or $b \rightarrow b_{\mathrm{m}}+0$ would be expressed by~\cite{Tsukamoto:2020uay,Tsukamoto:2020bjm,Zhang:2024sgs}
\begin{equation}\label{eq:def15}
\alpha 
= \frac{c_+}{ \left( \rho_0-\rho_{\mathrm{m}} \right)^{\frac{1}{2}}} +\bar{d}_+,
\end{equation}
where $c_+$ and $\bar{d}_+$ are constants. 
The coefficient $c_+$ and the term $\bar{d}_+$ satisfy the following relation 
\begin{equation}\label{eq:Eiroa11}
\lim_{\rho_0 \rightarrow \rho_{\mathrm{m}}+0}
\left[ \alpha-\frac{c_+}{ \left( \rho_0-\rho_{\mathrm{m}} \right)^{\frac{1}{2}}} -\bar{d}_+ \right] =0,
\end{equation}
where 
$\alpha$ is given by Eq.~(\ref{eq:deflection_angle1}),
and its derivative with respective to $\rho_0$ is given by 
\begin{equation}\label{eq:Eiroa12}
\lim_{\rho_0 \rightarrow \rho_{\mathrm{m}}+0}
\left[ \frac{\mathrm{d} \alpha}{\mathrm{d} \rho_0}+ \frac{c_+}{ 2 \left( \rho_0-\rho_{\mathrm{m}} \right)^{\frac{3}{2}}} \right] =0.
\end{equation}
From Eqs.~(\ref{eq:Eiroa11}) and (\ref{eq:Eiroa12}), we obtain $c_+$ and $\bar{d}_+$ numerically by calculating 
\begin{equation}\label{eq:Eiroa13}
c_+=
\lim_{\rho_0 \rightarrow \rho_{\mathrm{m}}+0}
\left[ -2  \left( \rho_0-\rho_{\mathrm{m}} \right)^{\frac{3}{2}} \frac{\mathrm{d} \alpha}{\mathrm{d} \rho_0} \right] 
\end{equation}
and
\begin{eqnarray}\label{eq:Eiroa14}
\bar{d}_+
&=&\lim_{\rho_0 \rightarrow \rho_{\mathrm{m}}+0}
\left[ \alpha-\frac{c_+}{ \left( \rho_0-\rho_{\mathrm{m}} \right)^{\frac{1}{2}}} \right] \nonumber\\
&=&\lim_{\rho_0 \rightarrow \rho_{\mathrm{m}}+0}
\left[ \alpha + 2\left( \rho_0-\rho_{\mathrm{m}} \right) \frac{\mathrm{d} \alpha}{\mathrm{d} \rho_0} \right].
\end{eqnarray}
Due to 
\begin{equation}
\left. \frac{\mathrm{d}b}{\mathrm{d}\rho_0} \right|_{\rho_0=\rho_{\mathrm{m}}}=0,
\end{equation}
the series of the expansion of $b(\rho_0)$ around $\rho_0=\rho_{\mathrm{m}}$ is obtained as 
\begin{equation}
b(\rho_0)=b_{\mathrm{m}}+ \frac{1}{2} \left. \frac{\mathrm{d}^2b}{\mathrm{d}\rho_0^2} \right|_{\rho_0=\rho_{\mathrm{m}}} \left( \rho_0 -\rho_{\mathrm{m}} \right)^2 + O\left(  \left( \rho_0 -\rho_{\mathrm{m}} \right)^3 \right),
\end{equation}
and the deflection angle~(\ref{eq:def15}) can be expressed by
\begin{equation}\label{eq:Eiroa19}
\alpha(b)= \frac{\bar{c}_+}{\left(\frac{b}{b_{\mathrm{m}}}-1\right)^\frac{1}{4} } +\bar{d}_+,
\end{equation}
where $\bar{c}_+$ is a constant given by 
\begin{equation}\label{eq:Eiroa20}
\bar{c}_+\equiv \left( \frac{1}{2b_{\mathrm{m}}} \left. \frac{\mathrm{d}^2b}{\mathrm{d}\rho_0^2} \right|_{\rho_0=\rho_{\mathrm{m}}} \right)^\frac{1}{4} c_+.
\end{equation}

\subsection{Rays passing through the throat}
The ray with the impact parameter $\left| b \right| < b_\mathrm{m}$ passes through the throat 
and it does not have the closest distance $r_0$ satisfying Eqs.~(\ref{eq:closest1}) and (\ref{eq:trajectory3}).
We use the deflection angle $\alpha (b)$ which is introduced by Shaikh~\textit{et al.}~\cite{Shaikh:2019jfr} in the following form:\footnote{
We comment on the definition of the deflection angle~(\ref{eq:deflection_angle2}).
One may consider that we should remove $-\pi$ from Eq.~(\ref{eq:deflection_angle2}) since it does not have much physical significance in the case of the passing through the throat.
We admit that it can be a valid point but we leave it as it is on this paper.
We note that the deflection angle is not an observable, we obtain observables by inserting the deflection angle into a lens equation,
and that we have some choices of the definitions of the deflection angle and the lens equation on the term $-\pi$.
We have chosen our expression of Eq.~(\ref{eq:deflection_angle2}) which is similar to Eq.~(\ref{eq:deflection_angle1}) 
since it is useful to treat a lens equation~(\ref{eq:lens}) with Eqs.~(\ref{eq:deflection_angle1}) and (\ref{eq:deflection_angle2}) at once in our lens configurations in Sec.~III
and since Eqs.~(\ref{eq:deflection_angle1}) and (\ref{eq:deflection_angle2}) are useful to compare with Shaikh~\textit{et al.}~\cite{Shaikh:2019jfr}.
See Perlick~\cite{Perlick:2003vg} for the definition of a lens equation and a deflection angle without $-\pi$.
}
\begin{eqnarray}\label{eq:deflection_angle2}
\alpha(b) 
&\equiv& 2 \int^\infty_{r_\mathrm{th}} \frac{\mathrm{d}r}{\sqrt{\frac{C(r)}{B(r)}\left( \frac{C(r)}{A(r)b^2} -1 \right)}} -\pi \nonumber\\
&=& 2 \int^\infty_{r_\mathrm{th}} \frac{ \left| b \right| \mathrm{d}r}{C(r) \sqrt{-V(r,b)}} -\pi.
\end{eqnarray}

We can expect that the deflection angle of rays inside of the marginally unstable photon sphere on the throat in the strong deflection limit $b \rightarrow b_{\mathrm{m}}-0$
has a similar form 
\begin{eqnarray}\label{eq:expectform}
\alpha(b)
&=& \frac{\bar{c}_-}{ \left| \frac{b}{b_{\mathrm{m}}}-1\right|^\frac{1}{4} } +\bar{d}_- \nonumber\\
&=& \frac{\bar{c}_-}{ \left( 1-\frac{b}{b_{\mathrm{m}}} \right)^\frac{1}{4} } +\bar{d}_-,
\end{eqnarray}
where $\bar{c}_-$ and $\bar{d}_-$ are constants,
to the deflection angle~(\ref{eq:Eiroa19}) in the strong deflection limit $b \rightarrow b_{\mathrm{m}}+0$
since the effective potential $V(r,b)$ has symmetry against the marginally unstable photon sphere on the throat.
The constants $\bar{c}_-$ and $\bar{d}_-$ are obtained from the following relation 
\begin{equation}\label{eq:Eiroa31}
\lim_{b \rightarrow b_{\mathrm{m}}-0}
\left[ \alpha-\frac{\bar{c}_- b_{\mathrm{m}}^\frac{1}{4}}{ \left( b_{\mathrm{m}}-b \right)^\frac{1}{4} } -\bar{d}_- \right] =0,
\end{equation}
where 
$\alpha$ is given by Eq.~(\ref{eq:deflection_angle2}),
and its derivative with respective to $b$ is given by 
\begin{eqnarray}\label{eq:Eiroa32}
\lim_{b \rightarrow b_{\mathrm{m}}-0}
\left[ \frac{\mathrm{d} \alpha}{\mathrm{d} b}- \frac{\bar{c}_- b_{\mathrm{m}}^\frac{1}{4} }{ 4   \left( b_{\mathrm{m}}-b \right)^\frac{5}{4} } \right] =0.
\end{eqnarray}
From Eqs.~(\ref{eq:Eiroa31}) and (\ref{eq:Eiroa32}), we obtain $\bar{c}_-$ and $\bar{d}_-$ numerically by calculating 
\begin{equation}\label{eq:Eiroa33}
\bar{c}_-=
\lim_{b \rightarrow b_\mathrm{m} -0}
\left[ \frac{4}{b_\mathrm{m}^\frac{1}{4}} \left( b_\mathrm{m}-b \right)^\frac{5}{4} \frac{\mathrm{d} \alpha}{\mathrm{d}b} \right] 
\end{equation}
and
\begin{eqnarray}\label{eq:Eiroa34}
\bar{d}_-
&=&\lim_{b \rightarrow b_\mathrm{m}-0}
\left[ \alpha-\frac{\bar{c}_- b_{\mathrm{m}}^\frac{1}{4}}{ \left( b_{\mathrm{m}}-b \right)^\frac{1}{4} } \right] \nonumber\\
&=&\lim_{b \rightarrow b_\mathrm{m}-0}
\left[ \alpha -4 \left( b_\mathrm{m} - b \right) \frac{\mathrm{d} \alpha}{\mathrm{d} b} \right].
\end{eqnarray}

\section{Gravitational lensing in the strong deflection limits}
We consider that a source S with a source angle $\beta \equiv \angle \mathrm{OLS}$ emits a ray with an impact parameter $b$ in the strong deflection limits $b \rightarrow b_{\mathrm{m}} \pm 0$, 
a lens L or a marginally unstable photon sphere on a throat reflects the ray with a deflection angle $\alpha$, 
and an observer O sees the source S as an image I with an image angle $\theta$. 
The lens configuration is shown as Fig.~1.
\begin{figure}[htbp]
\begin{center}
\includegraphics[width=80mm]{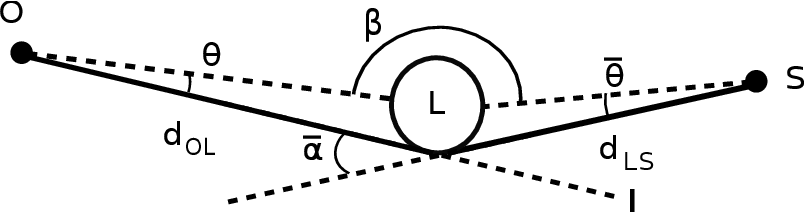}
\end{center}
\caption{A lens configuration with an effective deflection angle $\bar{\alpha}$, an image angle $\theta$, an angle $\bar{\theta}$, and a source angle $\beta$ is shown.
A source S emits a ray, the marginally unstable photon sphere of a wormhole at L reflects it, and an observer O observes it as an image I.   
Angular distances between O and L and between L and S are denoted by $d_{\mathrm{OL}}$ and $d_{\mathrm{LS}}$, respectively. 
We consider the cases that the source S is on the same and different regions of the observer O against the throat of the wormhole.}
\end{figure} 
The deflection angle $\alpha$ can be expressed by 
\begin{eqnarray}\label{eq:baralpha}
\alpha=\bar{\alpha}+ 2 \pi n,
\end{eqnarray}
where $\bar{\alpha}$ is the effective deflection angle of the ray defined by
\begin{eqnarray}
\bar{\alpha} \equiv \alpha \quad \mathrm{mod} \;  2 \pi
\end{eqnarray}
and a positive integer $n$ is the winding number of the ray. 
We define a new source angle $\gamma$ by
\begin{eqnarray}
\gamma \equiv \pi - \beta.
\end{eqnarray}
We use an Ohanian lens equation~\cite{Ohanian_1987,Bozza:2004kq,Bozza:2008ev} 
expressed by
\begin{eqnarray}\label{eq:lens}
\gamma = \bar{\alpha} -\theta -\bar{\theta},
\end{eqnarray}
where $\bar{\theta}$ is an angle between the ray and a line LS at the source S.

We assume that the angles $\left| \bar{\alpha} \right|$, $\left| \theta \right|$, $\left| \bar{\theta} \right|$, 
and $\left| \gamma \right|$ are small, i.e., $\left| \bar{\alpha} \right|$, $\left| \theta \right|$, $\left| \bar{\theta} \right|$, and $\left| \gamma \right| \ll 1$.
The image angle $\theta$ and the angle $\bar{\theta}$ can be expressed by 
\begin{eqnarray}
\theta= \frac{b}{d_{\mathrm{OL}}},
\end{eqnarray}
and 
\begin{eqnarray}
\bar{\theta}=\frac{b}{d_{\mathrm{LS}}},
\end{eqnarray}
respectively, where
$d_{\mathrm{OL}}$ and $d_{\mathrm{LS}}$ are angular distances 
between O and L and between L and S, respectively.
For simplicity, we ignore the angle $\bar{\theta}$ in the Ohanian lens equation.

For the marginally unstable photon sphere on the wormhole throat with $Z_2$ symmetry,
we can express the deflection angle $\alpha$ 
of the ray reflected just outside and inside of the marginally unstable photon sphere and its derivative respective to the image angle $\theta$ as
\begin{equation}\label{eq:alpha}
\alpha(\theta)=\frac{\bar{c}_\pm}{\left| \frac{\theta}{\theta_\infty}-1 \right|^{1/4}} +\bar{d}_\pm, 
\end{equation}
and 
\begin{equation}\label{eq:dalphadtheta}
\frac{\mathrm{d}\alpha}{\mathrm{d}\theta}=\mp \frac{\bar{c}_\pm}{ 4 \theta_\infty  \left| \frac{\theta}{\theta_\infty}-1 \right|^{5/4}}, 
\end{equation}
respectively,
where $\theta_\infty$ is an image angle for the marginally unstable photon sphere on the throat defined by 
\begin{equation}
\theta_\infty \equiv \frac{b_{\mathrm{m}}}{d_{\mathrm{OL}}}.
\end{equation}
We define $\theta^0_{n\pm}$ by 
\begin{eqnarray}\label{eq:theta0n}
\alpha \left(\theta^0_{n\pm} \right)=2\pi n
\end{eqnarray}
for $n\geq 1$.
By substituting it into Eq.~(\ref{eq:alpha}), 
we obtain $\theta^0_{n\pm}$ as 
\begin{eqnarray}\label{eq:theta0n2}
\theta^0_{n\pm} = \left[ 1 \pm \left( \frac{\bar{c}_\pm}{2\pi n - \bar{d}_\pm} \right)^4 \right] \theta_\infty.
\end{eqnarray}

By expanding the deflection angle $\alpha\left(\theta \right)$ around $\theta=\theta^0_{n\pm}$ for $n \geq 1$ as 
\begin{eqnarray}\label{eq:alphaexpand}
\alpha(\theta) 
&=&\alpha \left(\theta^0_{n\pm} \right)+\left.\frac{\mathrm{d}\alpha}{\mathrm{d}\theta}\right|_{\theta=\theta^0_{n\pm}} \left(\theta-\theta^0_{n\pm} \right) \nonumber\\
&&+O\left( \left( \theta-\theta^0_{n\pm} \right)^2 \right)
\end{eqnarray} 
and using Eqs. (\ref{eq:baralpha}),  (\ref{eq:dalphadtheta}), (\ref{eq:theta0n}), and (\ref{eq:alphaexpand}), 
we get the effective deflection angle as
\begin{eqnarray}\label{eq:baralpha2}
\bar{\alpha}\left(\theta_{n\pm} \right)=\mp \frac{\bar{c}_\pm \left( \theta_{n\pm} -\theta^0_{n\pm} \right) }{4 \theta_\infty \left| \frac{\theta^0_{n\pm}}{\theta_\infty} -1 \right|^{5/4}},
\end{eqnarray}
where $\theta=\theta_{n\pm}$ is the image angle for $n \geq 1$.
From Eqs.~(\ref{eq:lens}), (\ref{eq:theta0n2}), and (\ref{eq:baralpha2}),
the image angle $\theta_{n\pm}(\gamma)$ is obtained as  
\begin{equation}
\theta_{n\pm}(\gamma)\sim \theta^0_{n\pm} \mp  \frac{4 \bar{c}_\pm^4}{ \left( 2 \pi n - \bar{d}_\pm \right)^{5}} \theta_\infty (\theta^0_{n\pm} +\gamma).
\end{equation}
By setting $\gamma=0$, we obtain the image angle of the Einstein ring for $n \geq 1$ as
\begin{equation}
\theta_{\mathrm{E}n\pm} \equiv \theta_{n\pm}(0)
\sim \theta^0_{n\pm} \left[1 \mp \frac{4 \bar{c}_\pm^4 }{ \left( 2 \pi n - \bar{d}_\pm \right)^{5}} \theta_\infty \right].
\end{equation}
The difference of $\theta_{1\pm}$ and $\theta_\infty$ is given by
\begin{equation}
\bar{s}_\pm \equiv \theta_{1\pm} - \theta_\infty \sim \theta^0_{1\pm} - \theta^0_\infty = \pm \left( \frac{\bar{c}_\pm}{2\pi - \bar{d}_\pm} \right)^4 \theta_\infty.
\end{equation}

The magnification of the image with $\theta_{n\pm}$ for $n \geq 1$ is given by~\cite{Bozza:2004kq,Tsukamoto:2025hbz}  
\begin{eqnarray}
\mu_{n\pm}(\gamma) 
&\sim& \frac{d_{\mathrm{OS}}^2}{d_{\mathrm{LS}}^2} \frac{\theta_{n\pm}}{\gamma}\frac{\mathrm{d}\theta_{n\pm}}{\mathrm{d}\gamma} \nonumber\\
&\sim& -\frac{d_{\mathrm{OS}}^2}{d_{\mathrm{LS}}^2} \frac{4 \bar{c}_\pm^4 \left[\bar{c}_\pm^4 \pm (2\pi n-\bar{d}_\pm)^4  \right]}{(2\pi n-\bar{d}_\pm)^{9}} \frac{\theta_\infty^2}{\gamma},
\end{eqnarray}
where $d_{\mathrm{OS}}=d_{\mathrm{OL}}+d_{\mathrm{LS}}$ is an angular distance between O and S.
The ratio of $\mu_{1\pm}$ to the sum of the magnifications of the other images with $n\geq 2$ is given by 
\begin{eqnarray}
\bar{r}_\pm \equiv \frac{\mu_{1\pm}}{\sum^\infty_{n=2} \mu_{n\pm}} \sim \frac{K_{1\pm}}{\sum^\infty_{n=2} K_{n\pm}},
\end{eqnarray}
where $K_{n\pm}$ for $n \geq 1$ is defined by  
\begin{eqnarray}
K_{n\pm}\equiv \frac{\bar{c}_\pm^4 \pm (2\pi n-\bar{d}_\pm)^4}{(2\pi n-\bar{d}_\pm)^{9}}.
\end{eqnarray}

\section{Application}

\subsection{A commonly-used simple wormhole}
A commonly-used simple wormhole was investigated in Refs.~\cite{Harko:2008vy,Bambi:2013jda,Bambi:2013nla,Akiyama:2019cqa,EventHorizonTelescope:2022xqj}.
We cannot use a usual strong-deflection-limit analysis~~\cite{Bozza:2002af,Eiroa:2002mk,Petters:2002fa,Eiroa:2003jf,Bozza:2004kq,Bozza:2005tg,Bozza:2006sn,Bozza:2006nm,Iyer:2006cn,Bozza:2007gt,Tsukamoto:2016qro,Ishihara:2016sfv,Tsukamoto:2016oca,Tsukamoto:2016zdu,Tsukamoto:2016jzh,Shaikh:2018oul,Shaikh:2019itn,Takizawa:2021gdp,Zhang:2024sgs,Igata:2025taz,Tsukamoto:2021lpm,Sasaki:2023rdf} 
since the wormhole has a marginally unstable photon sphere at a throat.  
The line element is given by, in coordinates $(t, \rho, \vartheta, \varphi )$, 
\begin{equation}
\mathrm{d}s^2
=-e^{-2M/\rho}\mathrm{d}t^2+\frac{\mathrm{d}\rho^2}{1-\frac{M}{\rho}} +\rho^2 (\mathrm{d}\vartheta^2+\sin^2 \vartheta \mathrm{d}\varphi^2),
\end{equation}
where $M$ is a positive constant and where the throat and the marginally unstable photon sphere are at $\rho=\rho_\mathrm{m}=M$.  
The radial coordinate $M \leq \rho < \infty$ can be translated into a radial coordinate $r$ defined by
\begin{eqnarray}
r
&\equiv& \pm \int^\rho_{M} \sqrt{\frac{\rho}{\rho-M}} \mathrm{d}\rho \nonumber\\
&=& \pm \left( \sqrt{\rho^2-M\rho} \right. \nonumber\\
&& \left. +\frac{M}{2} \log \frac{2\rho -M +2\sqrt{\rho^2-M\rho}}{M} \right)
\end{eqnarray}
in a domain $-\infty < r < \infty$ and the throat and the marginally unstable photon sphere are at $r=r_\mathrm{m}=0$.
By using the radial coordinate $r$, we obtain the functions of the metric as 
\begin{eqnarray}
&&A(r)=e^{-\frac{2M}{\rho(r)}},\nonumber\\
&&B(r)=1,\nonumber\\
&&C(r)=\rho(r)^2.
\end{eqnarray}
We note that the assumptions that $A(r)$, $B(r)$, and $C(r)$ are positive and finite in the domain of the radial coordinate $-\infty<r<\infty$ are satisfied.
We can confirm that it holds the assumptions 
\begin{eqnarray}
&&D_\mathrm{m}=D^{\prime}_\mathrm{m}=D^{\prime \prime}_\mathrm{m}=0, \nonumber\\
&&D^{\prime \prime \prime}_\mathrm{m}=\frac{3}{2M^4}>0
\end{eqnarray}
so that it has the marginally unstable photon sphere at the throat.
Figure~2 shows the effective potential $V(r/M,b)$.
\begin{figure}[htbp]
\begin{center}
\includegraphics[width=60mm]{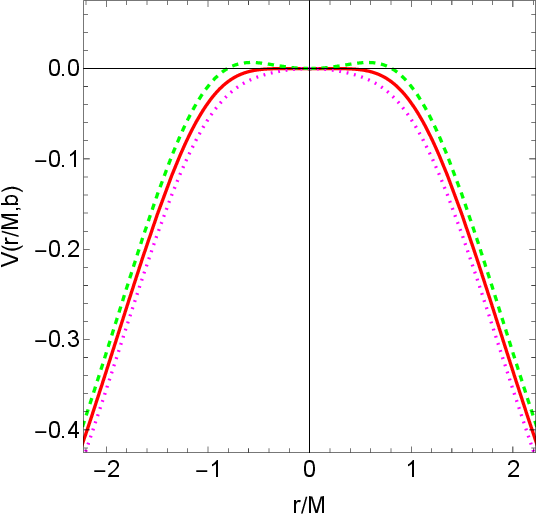}
\end{center}
\caption{The effective potential $V(r/M,b)$ of a commonly-used simple wormhole is shown.
The dotted~(magenta), solid~(red), and dashed~(green) curves denote $V(r/M,b)$ for $b=0.99 b_{\mathrm{m}}$, $b_{\mathrm{m}}$, and $1.01 b_{\mathrm{m}}$, respectively.}
\end{figure}
The impact parameter can be expanded as 
\begin{eqnarray}
b=b_\mathrm{m}+\frac{e}{2M} \left( \rho_0 -\rho_\mathrm{m} \right)^2 +O\left( \left( \rho_0 -\rho_\mathrm{m} \right)^3 \right),
\end{eqnarray}
where the critical impact parameter is given by 
\begin{eqnarray}
b_\mathrm{m}=M e.
\end{eqnarray}

From Eqs.~(\ref{eq:Eiroa33}) and (\ref{eq:Eiroa34}), we get
\begin{equation}\label{eq:cmCW1}
\bar{c}_- =6.24517
\end{equation}
and
\begin{equation}\label{eq:dmCW1}
\bar{d}_- =-3.93534
\end{equation}
and from Eqs.~(\ref{eq:Eiroa13}), (\ref{eq:Eiroa14}), and (\ref{eq:Eiroa20}), 
\begin{equation}\label{eq:cpCW1}
\bar{c}_+ =4.40973
\end{equation}
and
\begin{equation}\label{eq:dpCW1}
\bar{d}_+ =-3.6591.
\end{equation}
We define the percent errors of the deflection angle of Eq.~(\ref{eq:def02}) by,
for rays inside and outside of the marginally unstable photon sphere at the throat,  
\begin{equation}\label{eq:percentin}
\frac{\alpha_{(1.3)}-\alpha_{(2.39)}}{\alpha_{(2.39)}} \times 100,
\end{equation}
and 
\begin{equation}\label{eq:percentout}
\frac{\alpha_{(1.3)}-\alpha_{(2.29)}}{\alpha_{(2.29)}} \times 100
\end{equation}
respectively, where $\alpha_{(1.3)}$, $\alpha_{(2.29)}$, and $\alpha_{(2.39)}$ are the deflection angles in Eqs. $(1.3)$, $(2.29)$, and $(2.39)$, respectively. 
We show the deflection angles and its percent errors in Fig~3 
and summarize the coefficients, the terms, and observables in Table~I.
\begin{figure*}[htbp]
\begin{center}
\includegraphics[width=60mm]{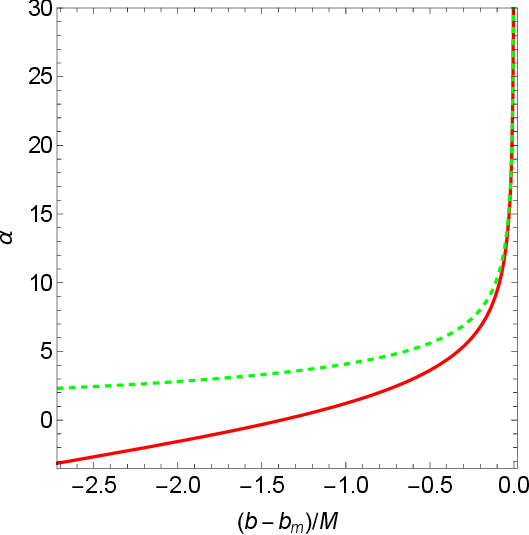}
\includegraphics[width=60mm]{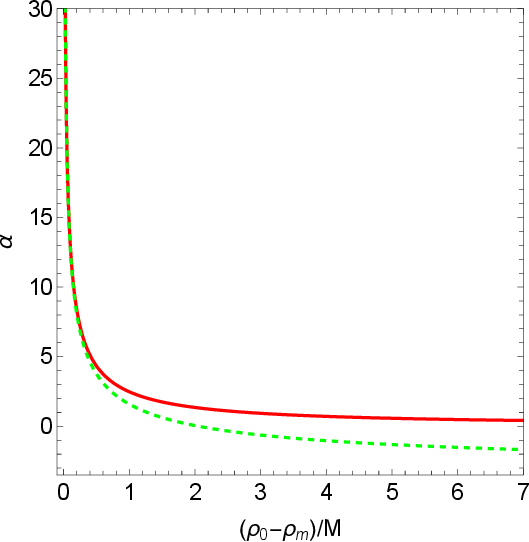}\\
\includegraphics[width=60mm]{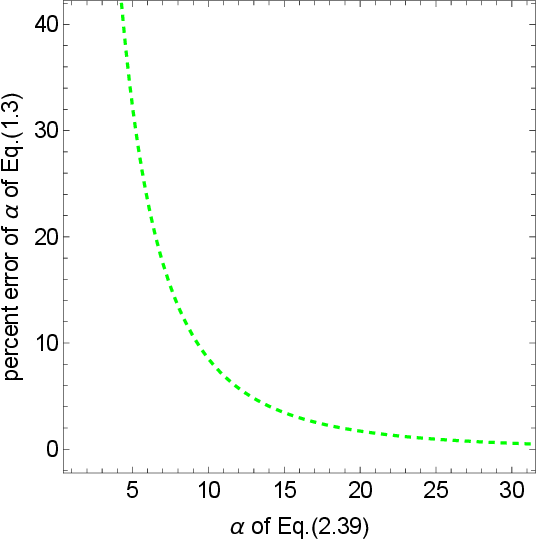}
\includegraphics[width=60mm]{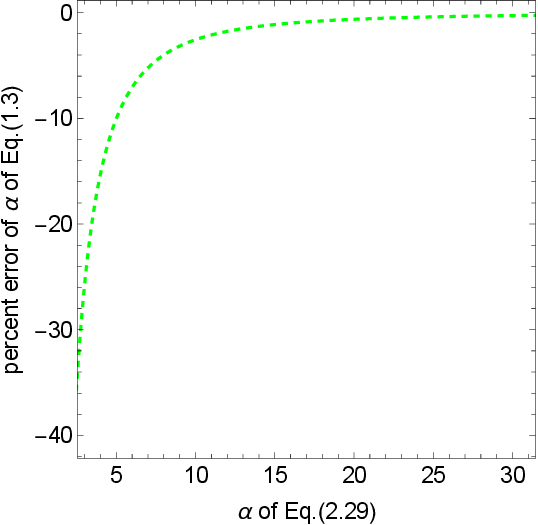}
\end{center}
\caption{Deflection angles by the commonly-used simple wormhole. 
Top-Left panel:Solid (red) and dashed (green) curves denote the deflection angles~$\alpha$ of Eqs.~(2.39) and (1.3), respectively,
inside of the marginally unstable photon sphere at the throat. 
Top-Right panel:Solid (red) and dashed (green) curves denote the deflection angle~$\alpha$ of Eqs.~(2.29) 
and (1.3), respectively,
outside of the marginally unstable photon sphere at the throat. 
Bottom-Left and bottom-right panels:The percent errors of the deflection angles~$\alpha$ of Eq.~(1.3),
against Eqs.~(2.39) and (2.29), respectively, are shown.
}
\end{figure*}
\begin{table*}[htbp]
 \label{table:I}
 \caption{In the commonly-used simple (CUS), Damour-Solodukhin (DS), and Simpson-Visser (SV) wormhole spacetimes with the marginally unstable photon sphere at the throat,
the critical impact parameter $b_\mathrm{m}$, 
the diameter of the marginally unstable photon sphere or the throat~$2\theta_{\infty}$, 
the coefficient $\bar{c}_\pm$ and the term $\bar{d}_\pm$ of the deflection angles in the strong deflection limits,  
the diameters of the Einstein rings $2\theta_{\mathrm{E} n \pm}$ with $n=1$, $2$, and $3$, 
the difference of the radii of the ring with $n=1$ and the marginally unstable photon sphere $\bar{s}_\pm=\theta_{1 \pm}-\theta_\infty$, 
the total magnifications of pair images $\mu_{\mathrm{tot}n \pm} \sim 2 \left| \mu_{n \pm} \right|$ for $n=1$, $2$, and $3$, and   
the ratio of the magnification of the ring with $n=1$ to the sum of the other rings $\bar{r}_\pm= \mu_{1 \pm}/\sum^\infty_{n=2} \mu_{n\pm}$ are shown.
We set $d_{\mathrm{OS}}=16$kpc, $d_{\mathrm{OL}}=d_{\mathrm{LS}}=8$kpc, $M=4\times 10^6 M_{\odot}$, and $\gamma=1$ arcsecond.}
\begin{center}
\begin{tabular}{c |c |c c c| c c} \hline
                                      &CUS       &DS        &DS\cite{Tsukamoto:2020uay} &DS\cite{Zhang:2024sgs} &SV        &SV~\cite{Tsukamoto:2020bjm} \\ \hline
 $b_\mathrm{m}/M$                     &$e$       &$3\sqrt{3}$ &$3\sqrt{3}$              &$3\sqrt{3}$            &$3\sqrt{3}$ &$3\sqrt{3}$                  \\
 $2\theta_{\infty}$ [$\mu$as]         &26.9833   &51.5802   &51.5802                    &51.5802        	&51.5802   &51.5802                    \\                           
 $\bar{c}_-/\bar{c}_+$                &1.4162    &1.4167    &$\cdots$                   &$\cdots$   		&1.4155    &$\cdots$                     \\ 
 $\bar{d}_-/\bar{d}_+$		      &1.0755    &1.1413    &$\cdots$                   &$\cdots$   	        &1.0515    &$\cdots$                     \\ \hline
 $\bar{c}_-$                          &6.2452    &4.7469    &$\cdots$                   &$\cdots$   	        &5.8087    &$\cdots$                     \\ 
 $\bar{d}_-$                          &-3.9353   &-2.4047   &$\cdots$                   &$\cdots$  	        &-3.9651   &$\cdots$                     \\ 
 $2\theta_{\mathrm{E}1-}$ [$\mu$as]   &23.2186   &46.9833   &$\cdots$                   &$\cdots$               &46.2567   &$\cdots$                     \\ 
 $2\theta_{\mathrm{E}2-}$ [$\mu$as]   &26.4298   &51.0589   &$\cdots$  		        &$\cdots$               &50.7939   &$\cdots$                     \\ 
 $2\theta_{\mathrm{E}3-}$ [$\mu$as]   &26.8310   &51.4518   &$\cdots$ 		        &$\cdots$               &51.3634   &$\cdots$                     \\ 
 $\bar{s}_-$ [$\mu$as]                &-1.88237  &-2.29846  &$\cdots$                   &$\cdots$               &-2.66173  &$\cdots$                     \\ 
 $\mu_{\mathrm{tot}1-}\times 10^{19}$ &3317.79   &9641.86   &$\cdots$   	        &$\cdots$               &9319.28   &$\cdots$                     \\ 
 $\mu_{\mathrm{tot}2-}\times 10^{19}$ &343.874   &689.596   &$\cdots$   	        &$\cdots$               &936.954   &$\cdots$                     \\ 
 $\mu_{\mathrm{tot}3-}\times 10^{19}$ &69.5587   &120.492   &$\cdots$   	        &$\cdots$               &189.255   &$\cdots$                     \\ 
 $\bar{r}_-$		              &7.38145   &11.1352   &$\cdots$   	        &$\cdots$               &7.61147   &$\cdots$                     \\ \hline
 $\bar{c}_+$                          &4.4097    &3.3507    &2.5558                     &2.5558   	        &4.1037    &3.1302                     \\ 
 $\bar{d}_+$                          &-3.6591   &-2.1069   &-2.1078                    &-2.1083   	        &-3.7709   &-2.7455                    \\ 
 $2\theta_{\mathrm{E}1+}$ [$\mu$as]   &28.0275   &52.8923   &52.0241   		        &52.0240                &53.0118   &52.3254                    \\ 
 $2\theta_{\mathrm{E}2+}$ [$\mu$as]   &27.1305   &51.7204   &51.6276   		        &51.6276                &51.7855   &51.6703                    \\ 
 $2\theta_{\mathrm{E}3+}$ [$\mu$as]   &27.0231   &51.6139   &51.5916   		        &51.5916                &51.6361   &51.6030                    \\ 
 $\bar{s}_+$ [$\mu$as]                &0.52210   &0.65604   &0.22198   	                &0.22192                &0.71579   &0.3726                     \\ 
 $\mu_{\mathrm{tot}1+}\times 10^{19}$ &1141.7    &3208.12   &1067.57   		        &1067.25                &2927.59   &1675.04                    \\ 
 $\mu_{\mathrm{tot}2+}\times 10^{19}$ &95.4703   &191.742   &64.7694   		        &64.7583                &252.442   &117.906                    \\ 
 $\mu_{\mathrm{tot}3+}\times 10^{19}$ &18.509    &32.2009   &10.8931   		        &10.8918                &49.4648   &21.1027                    \\ 
 $\bar{r}_+$		              &9.25652   &13.4520   &13.2478   		        &13.2459                &8.94947   &11.2411                    \\ \hline
\end{tabular} 
\end{center}
\end{table*}

\subsection{Damour-Solodukhin wormhole spacetime}
The Damour-Solodukhin wormhole spacetime~\cite{Damour:2007ap,Nandi:2018mzm,Ovgun:2018fnk,Bhattacharya:2018leh,Ovgun:2018swe} 
has the marginally unstable photon sphere on its throat at $\rho=\rho_\mathrm{m}= 3M$ when 
its line element is, in coordinates $(t, \rho, \vartheta, \varphi)$,
\begin{eqnarray}
\mathrm{d}s^2
&=&-\left(1-\frac{2M}{\rho}\right)\mathrm{d}t^2+\frac{\mathrm{d}\rho^2}{1-\frac{3M}{\rho}} \nonumber\\
&&+\rho^2 (\mathrm{d}\vartheta^2+\sin^2 \vartheta \mathrm{d}\varphi^2),
\end{eqnarray}
where $M$ is a positive constant.
The radial coordinate $\rho$ is defined in a domain $3M \leq  \rho < \infty$.
We transform the radial coordinate $\rho$ into a radial coordinate $r$ defined by
\begin{eqnarray}
r
&\equiv& \pm \int^\rho_{3M} \sqrt{\frac{\rho}{\rho-3M}} \mathrm{d}\rho \nonumber\\
&=& \pm \left( \sqrt{\rho^2-3M\rho} \right. \nonumber\\
&& \left. +\frac{3M}{2} \log \frac{2\rho -3M +2\sqrt{\rho^2-3M\rho}}{3M} \right)
\end{eqnarray}
in a domain $-\infty < r < \infty$ and the throat and the marginally unstable photon sphere are at $r=r_\mathrm{m}=0$.
By using the radial coordinate $r$, we obtain the functions of the metric as 
\begin{eqnarray}
&&A(r)=1-\frac{2M}{\rho(r)},\nonumber\\
&&B(r)=1,\nonumber\\
&&C(r)=\rho(r)^2.
\end{eqnarray}
We can confirm that there is the marginally unstable photon sphere at the throat $\rho=\rho_\mathrm{m}=3M$ or $r=r_\mathrm{m}=0$ due to
\begin{eqnarray}
&&D_\mathrm{m}=D^{\prime}_\mathrm{m}=D^{\prime \prime}_\mathrm{m}=0, \nonumber\\
&&D^{\prime \prime \prime}_\mathrm{m}=\frac{1}{18M^4}>0.
\end{eqnarray}
Figure~4 shows the effective potential $V(r/M,b)$. 
The impact parameter can be expanded as
\begin{eqnarray}
b=b_\mathrm{m}+\frac{\sqrt{3}}{2M} \left( \rho_0 -\rho_\mathrm{m} \right)^2 +O\left( \left( \rho_0 -\rho_\mathrm{m} \right)^3 \right),
\end{eqnarray}
where the critical impact parameter $b_\mathrm{m}$ is given by
\begin{eqnarray}
b_\mathrm{m}=3\sqrt{3} M.
\end{eqnarray}
\begin{figure}[htbp]
\begin{center}
\includegraphics[width=60mm]{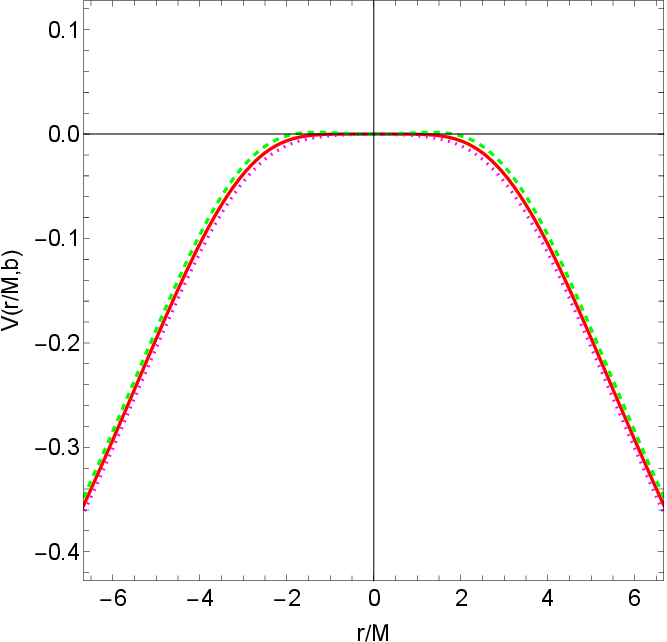}
\end{center}
\caption{The effective potential $V(r/M,b)$ with $b=0.99 b_{\mathrm{m}}$, $b_{\mathrm{m}}$, and $1.01 b_{\mathrm{m}}$ in the Damour-Solodukhin wormhole spacetime with the marginally unstable photon sphere are 
shown by the dotted~(magenta), solid~(red), and dashed~(green) curves, respectively.}
\end{figure} 

From Eqs.~(\ref{eq:Eiroa13}), (\ref{eq:Eiroa14}), and (\ref{eq:Eiroa20}), we obtain
\begin{equation}\label{eq:cmDS1}
\bar{c}_- =4.7469
\end{equation}
and
\begin{equation}\label{eq:dmDS1}
\bar{d}_- =-2.40469
\end{equation}
and from Eqs.~(\ref{eq:Eiroa13}), (\ref{eq:Eiroa14}), and (\ref{eq:Eiroa20}),  
\begin{equation}\label{eq:cpDS1}
\bar{c}_+ =3.3507
\end{equation}
and
\begin{equation}\label{eq:dpDS1}
\bar{d}_+ =-2.1069
\end{equation}
while 
Tsukamoto~\cite{Tsukamoto:2020uay} got the constants $\bar{c}_+$ and $\bar{d}_+$ as
\begin{equation}\label{eq:cpDS2}
\bar{c}_+=2.5558
\end{equation}
and
\begin{equation}\label{eq:dpDS2}
\bar{d}_+=-2.1078
\end{equation}
and Zhang and Xie~\cite{Zhang:2024sgs} got
\begin{equation}\label{eq:cpDS3}
\bar{c}_+=2.5558
\end{equation}
and
\begin{equation}\label{eq:dpDS3}
\bar{d}_+=-2.1083.
\end{equation}

Figure~5 and Table~I show that the semi-analytic calculations 
in Refs.~\cite{Tsukamoto:2020uay,Zhang:2024sgs} give invalid values of $\bar{c}_+$
while our method gives the correct value of the coefficient $\bar{c}_+$.
\begin{figure*}[htbp]
\begin{center}
\includegraphics[width=60mm]{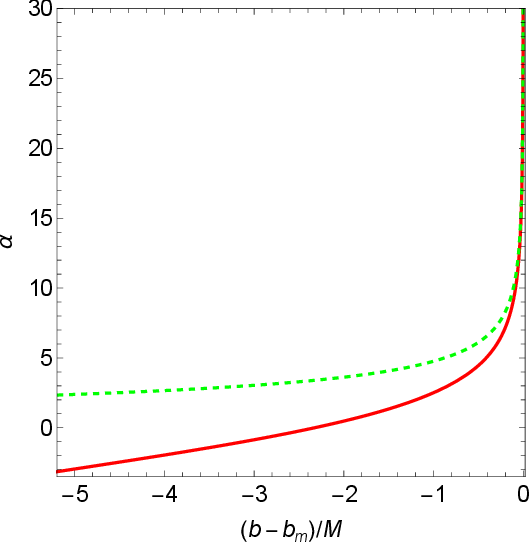}
\includegraphics[width=60mm]{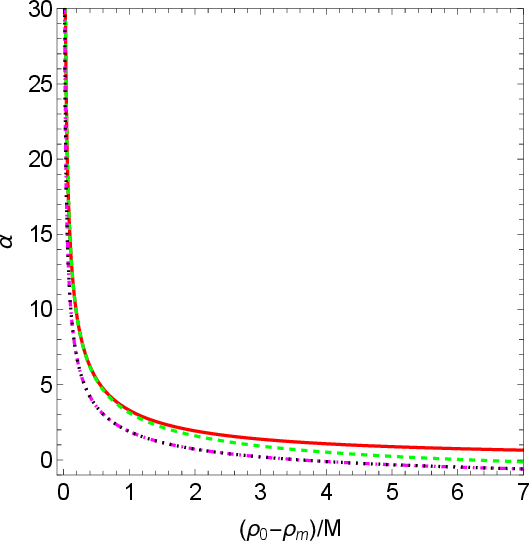}\\
\includegraphics[width=60mm]{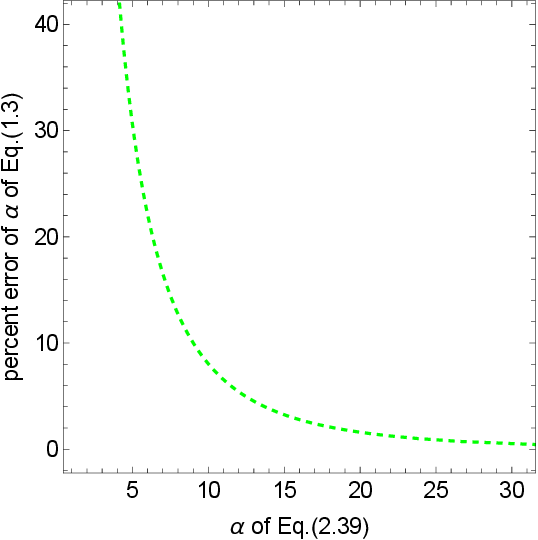}
\includegraphics[width=60mm]{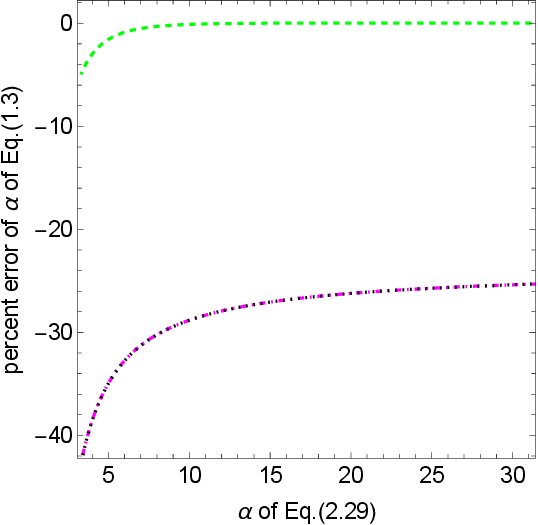}
\end{center}
\caption{
Deflection angles by the Damour-Solodukhin wormhole. 
Top-Left panel:Solid (red) and dashed (green) curves denote the deflection angles~$\alpha$ of Eqs.~(2.39) and (1.3), respectively,
inside of the marginally unstable photon sphere at the throat. 
Top-Right panel: 
A solid (red) curve denotes the deflection angle~$\alpha$ of Eqs.~(2.29), outside of the marginally unstable photon sphere at the throat. 
The deflection angles~$\alpha$~(1.3) with $\bar{c}_+$~(\ref{eq:cpDS1}) and $\bar{d}_+$~(\ref{eq:dpDS1}) in our method,   
with $\bar{c}_+$~(\ref{eq:cpDS2}) and $\bar{d}_+$~(\ref{eq:dpDS2}) by Tsukamoto~\cite{Tsukamoto:2020uay}, and
with $\bar{c}_+$~(\ref{eq:cpDS3}) and $\bar{d}_+$~(\ref{eq:dpDS3}) by Zhang and Xie~\cite{Zhang:2024sgs}
are denoted by dashed~(green), dot-dashed~(magenta), and black~(dotted) curves, respectively.
Note that the dot-dashed~(magenta) and black~(dotted) curves overlap.
Bottom-Left panel: The percent error of the deflection angle~$\alpha$ of Eq.~(1.3) against Eq.~(2.39) is shown.
Bottom-right panel: Dashed~(green), dot-dashed~(magenta), and black~(dotted) curves denote the percent errors of deflection angles~$\alpha$~(1.3)
with $\bar{c}_+$~(\ref{eq:cpDS1}) and $\bar{d}_+$~(\ref{eq:dpDS1}) in our method,   
with $\bar{c}_+$~(\ref{eq:cpDS2}) and $\bar{d}_+$~(\ref{eq:dpDS2}) by Tsukamoto~\cite{Tsukamoto:2020uay}, and
with $\bar{c}_+$~(\ref{eq:cpDS3}) and $\bar{d}_+$~(\ref{eq:dpDS3}) by Zhang and Xie~\cite{Zhang:2024sgs}, respectively,
against Eq.~(2.39).
}
\end{figure*}

\subsection{Simpson-Visser black-bounce spacetime}
The Simpson-Visser black-bounce spacetime describes a wormhole spacetime   
with the marginally unstable photon sphere on its throat at $l=l_\mathrm{m}=0$ when 
its line element is given by, in coordinates $(t, l, \vartheta, \varphi)$,
\begin{eqnarray}
\mathrm{d}s^2
&=&-\left(1-\frac{2M}{\sqrt{l^2+9M^2}}\right)\mathrm{d}t^2+\frac{\mathrm{d}l^2}{1-\frac{2M}{\sqrt{l^2+9M^2}}} \nonumber\\
&&+\left(l^2+9M^2 \right) (\mathrm{d}\vartheta^2+\sin^2 \vartheta \mathrm{d}\varphi^2),
\end{eqnarray}
where $M$ is a positive constant and the radial coordinate $l$ is defined in $-\infty<l<\infty$.
By transforming the radial coordinate $l$ into the radial coordinate $\rho$ given by
\begin{eqnarray}
\rho \equiv \sqrt{l^2+9M^2}
\end{eqnarray}
in a domain $\rho_\mathrm{m} \leq \rho < \infty $, where $\rho_\mathrm{m}=3M$,
the line element becomes
\begin{eqnarray}
\mathrm{d}s^2
&=&-\left(1-\frac{2M}{\rho}\right)\mathrm{d}t^2+\frac{\mathrm{d}\rho^2}{\left(1-\frac{9M^2}{\rho^2}\right) \left(1-\frac{2M}{\rho}\right)} \nonumber\\
&&+\rho^2 (\mathrm{d}\vartheta^2+\sin^2 \vartheta \mathrm{d}\varphi^2)
\end{eqnarray}
and the marginally unstable photon sphere and the throat are at $\rho=\rho_\mathrm{m}$.
By using the radial coordinate $r$, where $r$ is obtained as
\begin{eqnarray}
r
&\equiv& \pm \int^\rho_{3M} \sqrt{\frac{\rho^3}{\left(\rho^2-9M^2\right) \left(\rho-2M\right)}} \mathrm{d}\rho \nonumber\\
&=& \pm \left[ \sqrt{\frac{(\rho^2-9M)(\rho-2M)}{\rho}} \right. \nonumber\\
&&+\sqrt{\frac{3}{5}} \left( 3K \left( \sqrt{\frac{4}{5}} \right)   -5 E \left( \sqrt{\frac{4}{5}} \right) -2\Pi \left( 2, \sqrt{\frac{4}{5}} \right)   \right. \nonumber\\
&& \left. \left.  -3 F\left( \Phi, \sqrt{\frac{4}{5}} \right) +5 E \left( \Phi, \sqrt{\frac{4}{5}} \right) +2\Pi \left( 2, \Phi, \sqrt{\frac{4}{5}} \right)  \right) \right], \nonumber\\
\end{eqnarray}
where 
$F(\phi, k)$, $E(\phi, k)$, and $\Pi(n, \phi, k)$ 
are the incomplete elliptic integrals of the first, second, and third kinds defined by 
\begin{equation}
F(\phi, k) \equiv \int^\phi_0 \frac{\mathrm{d}\theta}{\sqrt{1-k^2 \sin^2 \theta}},
\end{equation}
\begin{equation}
E(\phi, k) \equiv \int^\phi_0 \sqrt{1-k^2 \sin^2 \theta} \mathrm{d}\theta,
\end{equation}
and
\begin{equation}
\Pi(n, \phi, k) \equiv \int^\phi_0 \frac{\mathrm{d}\theta}{\left( 1-n \sin^2 \theta \right) \sqrt{1-k^2 \sin^2 \theta}},
\end{equation}
respectively, and 
$K(k)$, $E(k)$, and $\Pi(n, k)$ 
are the complete elliptic integrals of the first, second, and third kinds defined by 
$K(k) \equiv F\left( \frac{\pi}{2}, k \right)$, $E(k) \equiv E\left( \frac{\pi}{2}, k \right)$, and $\Pi(n, k) \equiv \Pi \left( n, \frac{\pi}{2}, k \right)$, 
respectively, and 
$\Phi$ is defined by
\begin{eqnarray}
\Phi \equiv \arcsin \left( \sqrt{\frac{3M+ \rho}{2 \rho}} \right),
\end{eqnarray}
we obtain the functions of the metric as 
\begin{eqnarray}
&&A(r)=1-\frac{2M}{\rho(r)},\nonumber\\
&&B(r)=1,\nonumber\\
&&C(r)=\rho(r)^2
\end{eqnarray}
and the metric holds the conditions 
\begin{eqnarray}
&&D_\mathrm{m}=D^{\prime}_\mathrm{m}=D^{\prime \prime}_\mathrm{m}=0, \nonumber\\
&&D^{\prime \prime \prime}_\mathrm{m}=\frac{2}{81M^4}>0
\end{eqnarray}
at the marginally unstable photon sphere and the throat are at $\rho=\rho_\mathrm{m}=3M$ or $r=r_\mathrm{m}=0$.
Figure~6 shows the effective potential $V(r/M,b)$.
\begin{figure}[htbp]
\begin{center}
\includegraphics[width=60mm]{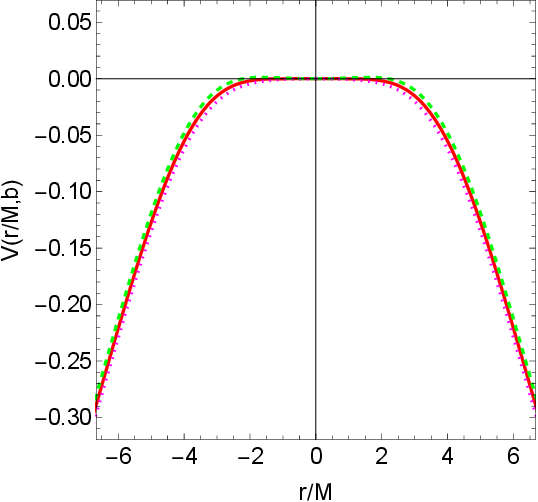}
\end{center}
\caption{The effective potential $V(r/M,b)$ in the Simpson-Visser black-bounce spacetime with the marginally unstable photon sphere is shown.
The dotted~(magenta), solid~(red), and dashed~(green) curves denote $V(r/M,b)$ for $b=0.99 b_{\mathrm{m}}$, $b_{\mathrm{m}}$, and $1.01 b_{\mathrm{m}}$, respectively.}
\end{figure}
The impact parameter can be expressed by 
\begin{eqnarray}
b=b_\mathrm{m}+\frac{\sqrt{3}}{2M} \left( \rho_0 -\rho_\mathrm{m} \right)^2 +O\left( \left( \rho_0 -\rho_\mathrm{m} \right)^3 \right),
\end{eqnarray}
and the critical impact parameter is given by
\begin{eqnarray}
b_\mathrm{m}=3\sqrt{3} M.
\end{eqnarray}

From Eqs.~(\ref{eq:Eiroa13}), (\ref{eq:Eiroa14}), and (\ref{eq:Eiroa20}), we obtain
\begin{equation}\label{eq:cmBB1}
\bar{c}_- =5.8087
\end{equation}
and
\begin{equation}\label{eq:dmBB1}
\bar{d}_- =-3.9651
\end{equation}
and, from Eqs.~(\ref{eq:Eiroa13}), (\ref{eq:Eiroa14}), and (\ref{eq:Eiroa20}), 
\begin{equation}\label{eq:cpBB1}
\bar{c}_+ =4.1037
\end{equation}
and
\begin{equation}\label{eq:dpBB1}
\bar{d}_+ =-3.7709
\end{equation}
while Tsukamoto~\cite{Tsukamoto:2020bjm} obtained 
\begin{equation}\label{eq:cpBB2}
\bar{c}_+= 3.13017
\end{equation}
and 
\begin{eqnarray}\label{eq:dpBB2}
\bar{d}_+= -2.74546.
\end{eqnarray}
From Fig~7, we have confirmed that our method gives the correct values of the coefficients $\bar{c}_\pm$ and the terms $\bar{d}_\pm$ 
but the invalid values of $\bar{c}_+$ and $\bar{d}_+$ were obtained in the semi-analytic calculation in Ref.~\cite{Tsukamoto:2020bjm}. 
\begin{figure*}[htbp]
\begin{center}
\includegraphics[width=60mm]{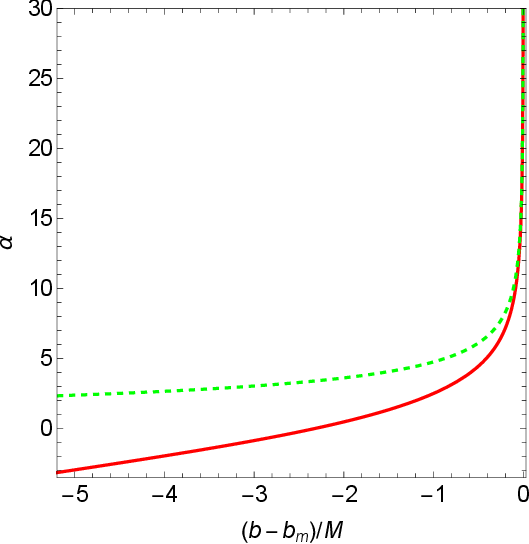}
\includegraphics[width=60mm]{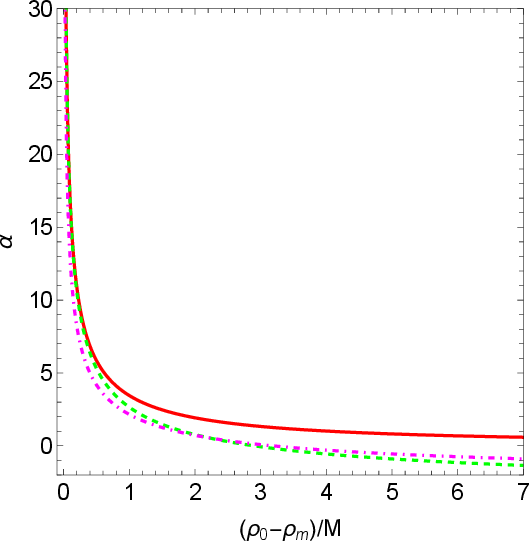}\\
\includegraphics[width=60mm]{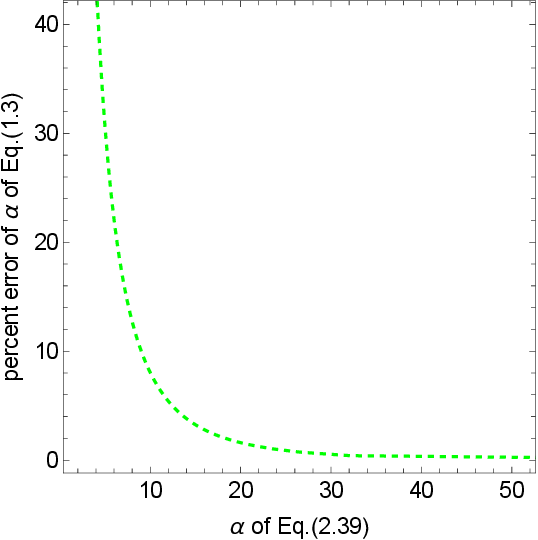}
\includegraphics[width=60mm]{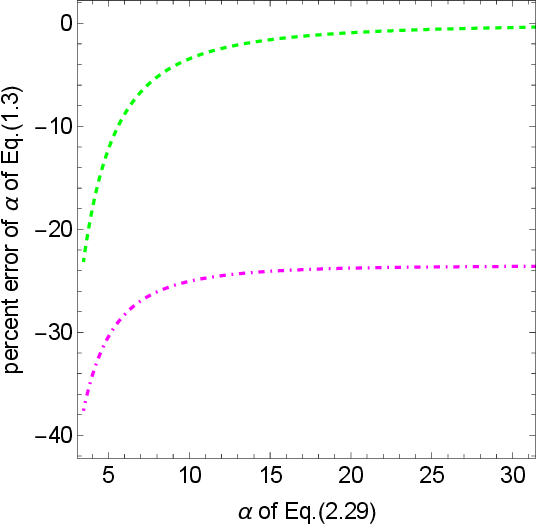}
\end{center}
\caption{
Deflection angles in the Simpson-Visser spacetime. 
Top-Left panel:Solid (red) and dashed (green) curves denote the deflection angle~$\alpha$ of Eqs.~(2.39) and (1.3), respectively,
inside of the marginally unstable photon sphere at the throat. 
Top-Right panel:
A solid (red) curve denotes the deflection angle~$\alpha$ of Eqs.~(2.29), outside of the marginally unstable photon sphere at the throat. 
The deflection angles~$\alpha$~(1.3) with $\bar{c}_+$~(\ref{eq:cpBB1}) and $\bar{d}_+$~(\ref{eq:dpBB1}) in our method,   
with $\bar{c}_+$~(\ref{eq:cpBB2}) and $\bar{d}_+$~(\ref{eq:dpBB2}) by Tsukamoto~\cite{Tsukamoto:2020bjm}
are denoted by dashed~(green), and dot-dashed~(magenta) curves, respectively.
Bottom-Left panel: The percent error of the deflection angle~$\alpha$ of Eq.~(1.3) against Eq.~(2.39) is shown.
Bottom-right panel: Dashed~(green) and dot-dashed~(magenta) curves denote the percent errors of deflection angles~$\alpha$~(1.3)
with $\bar{c}_+$~(\ref{eq:cpBB1}) and $\bar{d}_+$~(\ref{eq:dpBB1}) in our method, and  
with $\bar{c}_+$~(\ref{eq:cpBB2}) and $\bar{d}_+$~(\ref{eq:dpBB2}) by Tsukamoto~\cite{Tsukamoto:2020bjm}, respectively,
against Eq.~(2.39).
}
\end{figure*}

\section{Conclusion and Discussion}
The accretion disks and shadows in the commonly-uesd simple $Z_2$-symmetric wormhole~\cite{Harko:2008vy,Bambi:2013jda,Bambi:2013nla,Akiyama:2019cqa,EventHorizonTelescope:2022xqj} 
were investigated since its metric is apparently simple. 
We, however, have pointed out that the  wormhole has the marginally unstable photon sphere at the throat
and that it causes neither common nor simple but a special property that the deflection angle of the rays diverges in power law but not logarithmically 
near the marginally unstable photon sphere.
Therefore, we cannot apply a usual strong-deflection-limit analysis~\cite{Bozza:2002af,Eiroa:2002mk,Petters:2002fa,Eiroa:2003jf,Bozza:2004kq,Bozza:2005tg,Bozza:2006sn,Bozza:2006nm,Iyer:2006cn,Bozza:2007gt,Tsukamoto:2016qro,Ishihara:2016sfv,Tsukamoto:2016oca,Tsukamoto:2016zdu,Tsukamoto:2016jzh,Shaikh:2018oul,Shaikh:2019itn,Takizawa:2021gdp,Zhang:2024sgs,Igata:2025taz,Tsukamoto:2021fsz,Tsukamoto:2021lpm,Sasaki:2023rdf} to the wormhole.  

In this paper,  
We have investigated a strong-deflection-limit analysis of the rays with the deflection angles diverging
in power law outside and inside of the marginally unstable photon sphere at the $Z_2$-symmetric wormhole throat by extending Eiroa, Romero, and Torres' approch~\cite{Eiroa:2002mk} 
in numerical.
We apply our numerical method to the commonly-used simple, the Damour-Solodukhin, and the Simpson-Visser wormhole spacetimes.
Table~I implies that universal relations $\bar{c}_- = \sqrt{2} \bar{c}_+$ and $\bar{d}_- = \bar{d}_+$ between the coefficients $\bar{c}_\pm$ and the terms $\bar{d}_\pm$ 
in the deflection angles~(\ref{eq:def02}) of the marginally unstable photon sphere at the throat in the strong deflection limits~$b \rightarrow b_{\mathrm{m}} \pm 0$
while $\bar{c}_- = \sqrt{3} \bar{c}_+$ and $\bar{d}_- = \bar{d}_+$ in the deflection angles~(\ref{eq:def01}) 
of the marginally unstable photon sphere off the throat in the strong deflection limits~$b \rightarrow b_{\mathrm{m}} \pm 0$ obtained in analytical by Igata, Sasaki, and Tsukamoto \cite{Igata:2026ivq}. 
The difference of the ratio $\bar{c}_-$ to $\bar{c}_+$ will be caused by the different powers in Eqs.~(\ref{eq:def01}) and (\ref{eq:def02}).   

We have shown, from Fig.~5 and Table~I, that the coefficients $\bar{c}_+$ are 
invalid while the terms $\bar{d}_+$ is correct in the Damour-Solodukhin wormhole spacetime 
obtained in the semi-analytic method by Tsukamoto \cite{Tsukamoto:2020uay} and by Zhang and Xie~\cite{Zhang:2024sgs}.
From Fig.~7 and Table~I, we notice that both of $\bar{c}_+$ and $\bar{d}_+$ obtained in Ref.~\cite{Tsukamoto:2020bjm} are not valid in the Simpson-Visser wormhole spacetime.
Our results implies that the mismatch of the value of $\bar{d}_+$ in Ref.~\cite{Tsukamoto:2020bjm} will be likely due to simple calculation errors.

We have to mention that we have assumed that the deflection angles of the rays reflected near outside and inside of the marginally unstable photon sphere at the $Z_2$-symmetrical wormhole throat 
has the forms of Eqs.~(\ref{eq:def15}) and (\ref{eq:expectform}), respectively, in the strong deflection limits 
since the forms would be expected from earlier works~\cite{Tsukamoto:2020uay,Tsukamoto:2020bjm,Zhang:2024sgs}.
We should confirm that the forms~(\ref{eq:def15}) and (\ref{eq:expectform}) are correct in analytic calculations 
as well as Sasaki~\cite{Sasaki:2025web} and Igata, Sasaki, and Tsukamoto \cite{Igata:2026ivq} showed that 
the deflection angles of the rays reflected near outside and inside of the marginally unstable photon sphere off the throat 
have the form of Eq.~(\ref{eq:def01}) in the strong deflection limits in analytic calculations.  
It is left as a future work but it might not be completed by a simple extension of Ref.~\cite{Igata:2026ivq} 
since the deflection angle of the rays passing through the wormhole throat defined by Eq.~(\ref{eq:deflection_angle2}) is not the same as the usual deflection angle
defined by Eq.~(\ref{eq:deflection_angle1}).

%
\appendix

\end{document}